\documentclass[preprint,showpacs,preprintnumbers,amsmath,amssymb]{revtex4}

\usepackage{color}
\usepackage{graphicx}% Include figure files
\usepackage{dcolumn}% Align table columns on decimal point
\usepackage{bm}% bold math
\newcommand{\ft}[1]{\hat{#1}}

\begin{document}

\title{Density functional theory for the crystallization of two-dimensional dipolar colloidal alloys}

\author{W.R.C. Somerville$^\dag$, J.L. Stokes$^\dag$, A.M. Adawi$^\dag$, T.S. Horozov$^\dag$, A.J. Archer$^\ddag$ and D.M.A. Buzza$^{\dag,*}$}
 \affiliation{$^\dag$G. W. Gray Centre for Advanced Materials, School of Mathematics \& Physical Sciences, University of Hull, Hull HU6 7RX, UK, \\
 $^\ddag$Department of Mathematical Sciences, Loughborough University, Loughborough LE11 3TU, UK}

\date{\today}% It is always \today, today,

\begin{abstract}
Two-dimensional mixtures of dipolar colloidal particles with different dipole moments exhibit extremely rich self-assembly behaviour and are relevant to a wide range of experimental systems, including charged and super-paramagnetic colloids at liquid interfaces. However, there is a gap in our understanding of the crystallization of these systems because existing theories such as integral equation theory and lattice sum methods can only be used to study the high temperature fluid phase and the zero-temperature crystal phase, respectively. In this paper we bridge this gap by developing a density functional theory (DFT), valid at intermediate temperatures, in order to study the crystallization of one and two-component dipolar colloidal monolayers. The theory employs a series expansion of the excess Helmholtz free energy functional, truncated at second order in the density, and taking as input highly accurate bulk fluid direct correlation functions from simulation. Although truncating the free energy at second order means that we cannot determine the freezing point accurately, our approach allows us to calculate \emph{ab initio} both the density profiles of the different species and the symmetry of the final crystal structures. Our DFT predicts hexagonal crystal structures for one-component systems, and a variety of superlattice structures for two-component systems, including those with hexagonal and square symmetry, in excellent agreement with known results for these systems. The theory also provides new insights into the structure of two-component systems in the intermediate temperature regime where the small particles remain molten but the large particles are frozen on a regular lattice.
\end{abstract}

\pacs{82.70.Dd,68.65.Cd}% PACS, the Physics and Astronomy
\maketitle

\section{Introduction}
Colloidal monolayers at liquid interfaces have received a significant attention in the last two decades due to their importance both industrially and scientifically. From an industrial perspective, interfacial colloids are important in areas ranging from pharmaceuticals to food and personal care products \cite{Binks2002}. From a scientific perspective, interfacial colloids serve as ideal model systems to study self-assembly in two dimensional (2D) condensed matter systems due to the strong confinement of the colloidal particles by the liquid interface \cite{Bresme2007}.

A particularly important class of interactions for interfacial colloids are isotropic dipolar interactions. If we consider colloidal monolayers at an oil/water interface for example, the dipoles can arise from residual charges at the particle/oil interface together with their image charge in the water sub-phase \cite{Aveyard2000,Aveyard2002} or from the asymmetric electric double layer \cite{Stillinger1961,Hurd1985} or Stern layer \cite{Masschaele2010} at the particle/water interface. For all the aforementioned systems, the dipoles are electric dipoles. However, such dipolar interactions are also relevant for colloids in other confined geometries, e.g., super-paramagnetic particles confined by gravity to be on a flat air/water interface with a magnetic field applied perpendicular to the liquid interface, leading to magnetic dipoles which are perpendicular to the interface \cite{Zahn1999}.

In recent years, there has been considerable interest in binary mixtures of dipolar colloids with different dipole moments at liquid interfaces due to the very rich self-assembly behaviour exhibited by these systems. For example, in the high temperature regime where the system exists in the fluid phase, the mixture is found to form a microphase separated structure where the smaller particles cluster around the larger particles \cite{Hoffmann2006,Hoffmann2006b}. On the other hand, in the low temperature regime, these systems form a bewildering variety of super-lattice crystal structures \cite{Stirner2005,Assoud2007,Fornleitner2009,Chremos2009,Law2011,Law2011b}. In contrast, the intermediate temperature regime where the system is close to the crystallization point is much less studied, though the limited studies that do exists suggest that the self-assembly behaviour here is equally rich \cite{Law2011b}.

This gap in our understanding is largely due to the lack of suitable theoretical tools for studying the intermediate temperature regime. Specifically, while integral equation theory can be used to study the fluid state at high temperature \cite{Hoffmann2006,Hoffmann2006b} and lattice sum methods for the crystal state at zero temperature \cite{Assoud2007,Fornleitner2009,Chremos2009,Law2011,Law2011b}, neither of these methods are applicable to the region around the crystllization point. The lattice sum method has the further limitation that it can only be used to study spatially periodic structures, thus excluding the possibility of studying non-periodic ordered structures such as quasi-crystals \cite{Scheffler2004}. Particle based simulations (e.g., Monte Carlo simulations) provide us with some possibilities in addressing this problem \cite{Stirner2005,Law2011b}. However, the well known limitations of this approach, such as fluctuations in the local densities, slow dynamics when exploring the rugged free energy landscape of complex systems etc., mean that it is not always possible to use this approach to obtain a comprehensive and reliable set of stable crystal structures.

The aim of this paper is to address this problem by developing a density functional theory (DFT) for the crystallization of binary dipolar colloids in two-dimensions. DFT is a powerful technique for studying the microscopic density distribution in condensed matter systems, including the crystalline state. Specifically, it asserts that the Helmholtz free energy of the system is a unique functional of the one-particle densities of the different species in the system \cite{Hansenbook, Evans1979, evans1992fundamentals}. Since DFT treats both the fluid and crystal state on the same footing, it allows us to study the freezing transition accurately. Specifically, it returns the \emph{average} local density of the different species in the system, thus providing accurate information about the crystal structure which is not obscured by noise. In addition, since DFT is based on free energy minimization rather than any underlying dynamics of the system, it is much faster and more reliable compared to particle-based simulation methods in finding stable crystal structures.

The first DFT for two dimensional dipolar systems was constructed by van Teeffelen \emph{et al.} \cite{vanTeeffelen2006,vanTeeffelen2008}. These authors performed a series expansion of the Helmholtz free energy in density fluctuations around a reference liquid state, effectively treating the crystal as a spatially inhomogeneous fluid. By using very accurate free energy functionals for the system, where higher than second order terms in the series expansion were included either perturbatively or non-perturbatively, these authors were able to accurately predict the freezing point for dipolar systems. However, van Teeffelen \emph{et al.} only considered the case of one-component dipolar systems. Furthermore, to make their calculations tractable, they used predefined forms for the colloid density profiles and performed their calculations over one unit cell with a predetermined symmetry. In this paper, we use a simpler free energy functional for the system, expanding the Helmholtz free energy functional only up to second order in the density. We further employ a heuristic scaling approximation to extrapolate the direct correlation function (i.e., the coefficient of the second order term) beyond the crystallization point and induce the system to crystallize. Although truncating the free energy at second order means that we are no longer able to determine the freezing point accurately, our approach allows us to lift a number of the constraints of the approach in Ref.~\cite{vanTeeffelen2006,vanTeeffelen2008}. Firstly, we are able to consider both one and two-component dipolar monolayers. Secondly, we do not need to make any \emph{a priori} assumptions concerning the density profiles of the particles; instead the density fields of the different components are returned as an output of our calculation. Finally, we are able to perform our calculations over large areas containing many unit cells so that the system is not constrained to have a specific crystal structure but is free to choose its optimum crystal structure. Our method therefore complements the approach of van Teeffelen \emph{et al.}, providing a simple but powerful \emph{predictive} tool for studying the crystallization of one- and two-component dipolar monolayers.

The rest of the paper is structured as follows. In section II we discuss the background theory, including details of system parameters, the density functional theory and the integral equation theory and Monte Carlo simulations on which the DFT is based. In section III we discuss results from our DFT for both one- and two-component dipolar systems and compare these to known results from other methods. Finally in section IV, we give our conclusions.

\section{Background Theory}
\subsection{The system and interaction parameters}\label{system}
We first consider the one-component system consisting of $N$ colloidal particles with radius $R$ at a flat liquid interface with area $A$. Each particle possesses a dipole (electric or magnetic) of magnitude $P$ which is oriented perpendicular to the interface. We use the typical distance between particles $a \equiv \rho^{-1/2}$ as our characteristic length scale, where $\rho =N/A$ is the two-dimensional number density of colloids. For low enough colloid densities such that $a>>2R$ (a condition that is easily satisfied experimentally for many dipolar systems of interest), we can neglect the short range hard core repulsion and treat each colloid as a point-like dipole. In this case, the interaction between two particles with centre-to-centre separation $r$ is given by
\begin{equation}\label{u1}
  \beta U(r) = \Gamma \frac{a^3}{r^3}
\end{equation}
where $\beta=1/k_BT$, $k_B$ is Boltzmann's constant, $T$ is the absolute temperature and $\Gamma$ is the dimensionless dipole interaction strength. Physically, $\Gamma$ measures the strength of the dipolar interaction between two particles which are at the typical separation $a$ relative to the thermal energy.

As mentioned in the Introduction, such dipolar interactions naturally arise in many 2D colloidal systems. For example, for colloids adsorbed at an oil/water interface with contact angle $\theta$, the dipoles can arise from residual charges at the particle/oil interface together with their image charge in the water sub-phase \cite{Aveyard2000,Aveyard2002}. In this case $\Gamma = P^2/8 \pi \epsilon_r \epsilon_0 a^3 k_BT$, where $P = 2 q \zeta$, $q= 2 \pi R^2 \sigma (1-\cos \theta)$, $\zeta = R (3+\cos \theta)/2$, $\sigma$ is the surface charge density at the particle/oil interface, $\epsilon_r$ is the relative permittivity of the oil phase and $\epsilon_0$ is the vacuum permittivity. On the other hand, for super-paramagnetic particles on an air/water interface, an induced magnetic dipole arises when a magnetic field $B$ is applied perpendicular to the interface. In this case $\Gamma = \mu_0 P^2/4 \pi a^3 k_BT$, where $\mu_0$ is the vacuum permeability, $P= \chi B$ and $\chi$ is the magnetic susceptibility of the particle as a whole \cite{Zahn1999}. Note that the thermodynamic state of a one-component dipolar system is fully characterised by the interaction strength $\Gamma$ and does not depend separately on temperature and density.

We can generalise the above to two-component dipolar systems. In this case, each component is characterised by its own density $\rho_1$, $\rho_2$, radius $R_1$, $R_2$ and dipole moment strength $P_1$, $P_2$. For definiteness, we define species $1$ as the species with the larger dipole moment and use the typical separation between these particles $a \equiv \rho_1^{-1/2}$ as the characteristic length scale for the two-component system. Since dipole moment strength generally correlates with particle size in the experimental system, we will also refer to species $1$ and $2$ as `large' and `small' respectively in what follows. For large enough separations $r$, the interaction between two particles of species $i$ and $j$ respectively ($i,j=1,2$) is given by
\begin{equation}\label{u2}
  \beta U_{ij}(r) = \Gamma m_i m_j \frac{a^3}{r^3}
\end{equation}
where $m_i=P_i/P_1$ is the dipole moment for species $i$ relative to that of a large particle. In Eq.~\eqref{u2}, $\Gamma$ is the dimensionless interaction strength between large particles. For example, for charged colloids at an oil/water interface, $\Gamma = P_1^2/8 \pi \epsilon_r \epsilon_0 a^3 k_BT$. Note that the thermodynamic state of a two-component dipolar system is fully characterised by three parameters: the interaction strength $\Gamma$, the dipole moment ratio $m_2$ and the density ratio $\rho_2/\rho_1$.

\subsection{Integral equation theory}\label{IET}
As mentioned in the Introduction, the DFT we implement uses the fluid state as the reference state and therefore requires an accurate description of the fluid state as its starting point. The latter is provided by integral equation theory. For a one-component system, integral equation theory describes the structure of the fluid via two correlation functions, namely the radial distribution function $g(r)$ and the two-body direct correlation function $c(r)$. Physically, $\rho g(r)$ corresponds to the density of colloids at a distance $r$ from the origin given that another colloid is located at the origin. These two functions are related to each other via the Ornstein-Zernicke (OZ) relation \cite{Hansenbook}
\begin{equation}\label{OZ}
  h(r)=c(r)+\rho \int d\mathbf{r}' c(|\mathbf{r}-\mathbf{r}'|) h(r')
\end{equation}
where $h(r) \equiv g(r)-1$ is called the total correlation function and the integral is carried out over 2D space for our system. For actual calculations, it is more convenient to express the OZ relation in Fourier space,
\begin{equation}\label{OZ2}
  \ft{h}(q) = \ft{c}(q) +\rho \ft{h}(q) \ft{c}(q)
\end{equation}
where $\ft{h}(q)$ and $\ft{c}(q)$ are the 2D Fourier transforms of $h(r)$ and $c(r)$ respectively. The OZ relation is exact, but since it connects two unknown functions, one more relation or \emph{closure} is needed in order to determine $h(r)$ and $c(r)$.

In order to obtain an accurate DFT, we require accurate expressions for $c(r)$ close to the freezing transition (see subsection \ref{DFT}). However, van Teeffelen \emph{et al.} have shown that even closures like the hypernetted chain (HNC) and Rogers-Young (RY), which are generally considered to be accurate for soft interactions such as dipolar interactions, are not accurate enough for this purpose \cite{vanTeeffelen2006,vanTeeffelen2008}. We therefore use Monte Carlo (MC) simulations instead to obtain high precision radial distribution functions $g_s(r)=h_s(r)+1$, where the subscript $s$ refers to quantities obtained from simulations (see next subsection for further details). Since the accessible range of $h_s(r)$ is limited to $r \leq L/2$, where $L$ is the simulation box size, we used the extrapolation method of Verlet \cite{Verlet1968} to extend the correlation functions to large enough $r$ so that Fourier transforms can be performed. Specifically, we used the following closure (the Verlet-HNC closure) to relate $h(r)$ and $c(r)$
\begin{eqnarray}
  h(r) &=& h_s(r), \; r<r_c \nonumber \\
  c(r) &=& c_{HNC}(r), \; r>r_c \label{verlet}
\end{eqnarray}
where $r_c$ is a suitably chosen cut-off radius for the simulation data such that $r_c \leq L/2$. In Eq.~\eqref{verlet}, $c_{HNC}(r)$ is the direct correlation function obtained from the HNC closure
\begin{equation}\label{HNC}
  c(r)=e^{-\beta U(r)+\gamma(r)}-\gamma(r)-1
\end{equation}
where $\gamma(r) \equiv h(r)-c(r)$ is the indirect correlation function. In practice, it is numerically more stable to work in terms of $\gamma (r)$ and $c(r)$ rather than $h(r)$ and $c(r)$. In this case, the Verlet-HNC closure can be rewritten as
\begin{equation}\label{verlet2}
  c(r) =    \begin{cases}
                h_s (r)-\gamma(r), & r \leq r_c \\
                c_{HNC}(r), & r>r_c
            \end{cases}
\end{equation}
The OZ relation and the Verlet-HNC closure can now be solved iteratively to obtain the correlation functions $\gamma(r)$ and $c(r)$.

The above discussion can be readily generalised to two-component systems. In this case, the fluid structure is described by integral equation theory using three total correlation functions $h_{ij}(r)$ and three direct correlation functions $c_{ij}(r)$ where $i,j=1,2$; due to symmetry $h_{12}(r)=h_{21}(r)$ and $c_{12}(r)=c_{21}(r)$. In our calculations, these functions are related to each other via the two-component OZ relations and Verlet-HNC closures. Specifically, using the Einstein summation convention, the two-component OZ relation in Fourier space is given by the matrix equation \cite{Hoffmann2006b}
\begin{equation}\label{OZ3}
  \ft{h}_{il}(q)=\ft{c}_{il}(q)+\ft{c}_{ij}(q) d_{jk} \ft{h}_{kl}(q)
\end{equation}
where $\ft{h}_{ij}(q)$, $\ft{c}_{ij}(q)$ are the 2D Fourier transforms of $h_{ij}(r)$, $c_{ij}(r)$ respectively, $d_{ij}$ is the diagonal matrix of partial densities
\begin{equation}\label{d}
  d_{ij} =\rho_i \delta_{ij},
\end{equation}
$\delta_{ij}$ is the Kronecker delta and no summation is implied by the repeated index $i$ above. On the other hand, the two-component Verlet-HNC closure is given by
\begin{equation}\label{verlet3}
  c_{ij}(r) =    \begin{cases}
                 h_{s,ij} (r)-\gamma_{ij}(r), & r\leq r_c \\
                 c_{HNC,ij}(r), & r>r_c
            \end{cases}
\end{equation}
where $h_{s,ij} (r)$ are the total correlation functions obtained from simulation, $c_{HNC,ij}(r)$ are the direct correlation functions obtained from the HNC closure
\begin{equation}\label{HNC2}
  c_{ij}(r)=e^{-\beta U_{ij}(r)+\gamma_{ij}(r)}-\gamma_{ij}(r)-1
\end{equation}
and $\gamma_{ij} (r) \equiv h_{ij}(r)-c_{ij}(r)$ are the indirect correlation functions of the two-component system. For practical calculations, the OZ relation given by Eqs.~\eqref{OZ3} and \eqref{d} and the Verlet-HNC closure relations given by Eqs.~\eqref{verlet3} and \eqref{HNC2} are solved iteratively in terms of $\gamma_{ij} (r)$ and $c_{ij} (r)$ rather than $h_{ij}(r)$ and $c_{ij}(r)$ as before. All Fourier transforms are performed using the Discrete Hankel transform \cite{Johnson1987,Lemoine1994}. This is equivalent to a 2D fast Fourier transform method where the function to be transformed has radial symmetry, but has the benefit of being expressed as a 1D transform. The order used was $2048$, which was found to have converged, and a maximum value of $r/a=30$ was used in the transforms, which is sufficiently large that the function has decayed enough to not impact the results.

Another important quantity used to describe the structure of the system is the structure factor. For a one-component system, the structure factor is given by
\begin{equation}\label{sq}
  S(q)=1+\rho \ft{h}(q),
\end{equation}
while for a two-component system, the three partial structure factors are given by \cite{Hoffmann2006b}
\begin{equation}\label{sq2}
  S_{ij}(q) = \delta_{ij}+\sqrt{\rho_i \rho_j} \ft{h}_{ij} (q).
\end{equation}
Here $\ft{h}(q)$ and $\ft{h}_{ij} (q)$ are the Fourier transforms of the total correlation functions $h(r)$ and $h_{ij}(r)$ respectively.

\subsection{Monte Carlo simulations}\label{MC}
As explained in the previous subsection, we use Monte Carlo simulations as an input to our integral equation theory in order to obtain accurate correlation functions. Verlet \cite{Verlet1968} showed that very accurate results for these functions could be obtained even using relatively small values of the cut-off radius $r_c$ in the Verlet-HNC closure (see Eqs.\ \eqref{verlet2} and \eqref{verlet3}). Practically, this means that we only need to perform simulations on relatively small systems in the fluid state, which are computationally very cheap. Specifically, for the one-component systems, we use $N=576$ particles in a $24a \times 24a$ square box with periodic boundary conditions (for different values of $\Gamma$) while for two-component systems we use $N=576$ large particles in a $24a \times 24a$ square box with periodic boundary conditions (for different values of $\Gamma$, $m_2$ and $\rho_2/\rho_1$). Unless stated otherwise, in all our simulations, the initial state (a random distribution of particles) was first equilibrated for $10^4$ MC steps per particle. After equilibration, $4\times10^4$ MC steps per particle are used for the analysis phase, with all quantities obtained by averaging over $4000$ snapshots, i.e., 1 snapshot for every 10 MC steps per particle to ensure the snapshots are independent. The maximum MC step length was adjusted to ensure an acceptance probability of around 30\% throughout the simulation.

In order for the correlation functions calculated from integral equation theory to be valid, we need to ensure that the simulated system remains in the fluid phase. On the other hand, in order to obtain an accurate DFT for the crystal phase, we require accurate results for $c(r)$ very close to the freezing transition. These two constraints mean that it is imperative that we determine the freezing point of our simulated system $\Gamma_c$ accurately so that we can work with $\Gamma$ values which are close to, but still below, $\Gamma_c$.

The freezing point $\Gamma_c$ was determined from our MC simulations in two ways. Firstly, it was determined by measuring the $n$-fold orientational order parameter of the system, which is defined as
\begin{equation}\label{oop}
  \Psi_n = \left\langle \left| \frac{1}{N} \sum_{k=1}^N \psi_{n,k} \right| \right\rangle,
\end{equation}
where $N$ is the total number of particles, $\langle ... \rangle$ denotes an average over MC snapshots, and $\psi_{n,k}$ is the local $n$-fold orientational order parameter around the $k$-th particle defined as
\begin{equation}\label{oop2}
  \psi_{n,k}=\frac{1}{N_k} \sum_{j=1}^{N_k} \exp(i n \theta_{kj}).
\end{equation}
Here the sum $j$ is over the $N_k$ nearest neighbours of particle $k$ (where nearest neighbours are those particles that share an edge in a Delauny triangulation), and $\theta_{kj}$ is the angle between $\mathbf{r}_{kj}$, the displacement vector from particle $k$ to $j$, and the $x$-axis. For two-component systems, $\Psi_n$ is defined as the orientational order parameter for the large particles only. The order parameter $\Psi_n$ depends very sensitively on the average orientational order of the system, for example $\Psi_6=0$ for a fluid while $\Psi_6=1$ for a perfect hexagonal crystal or hexatic phase (i.e., a 2D phase with orientational but not translational order). Therefore, provided the simulated system actually crystallizes, we can use this order parameter to determine $\Gamma_c$.

An important technical detail that we should mention here is the fact that the crystallization of one-component dipolar systems has been experimentally demonstrated to proceed via two stages, i.e., first a fluid to hexatic transition, then a hexatic to hexagonal crystal transition \cite{Zahn1999b}, consistent with the theoretical predictions of Kosterlitz, Thouless, Halperin, Nelson and Young (KTHNY) \cite{Kosterlitz1973,Halperin1978,Nelson1979,Young1979}. Strictly speaking therefore, the orientational order parameter $\Psi_n$ measures the fluid to hexatic phase transition point rather than the crystallization point per se. However, since the $\Gamma$ value for the fluid to hexatic transition is less than 10\% lower than that for the hexatic to crystal transition for dipolar systems \cite{Zahn1999b}, it is accurate enough for our purposes to use $\Psi_n$ to measure the crystallization point.

Although we can use the orientational order parameter to measure $\Gamma_c$, it is well known from experiments on binary dipolar systems \cite{Ebert2008} that for larger values of the dipole moment ratio $m_2$, the fluid phase can be arrested by a glass transition before crystallization can occur. To account for this possibility, we therefore also determine $\Gamma_c$ (in this context the point where the system ceases to be a fluid because of crystallization or a glass transition) by measuring the 2D dynamic Lindemann parameter which we define as \cite{Zheng1998,Zahn1999b,Parolini2015}
\begin{equation}\label{lindemann}
  \gamma_L(t) = \frac{1}{2 a^2}\left\langle \left[ \Delta \mathbf{r}_k(t)-\frac{1}{N_k}\sum_{j=1}^{N_k} \Delta \mathbf{r}_j(t) \right] \right\rangle.
\end{equation}
Here $\Delta \mathbf{r}_i(t) = \mathbf{r}_i(t)-\mathbf{r}_i(t=0)$ is the displacement of the $i$-th particle from an arbitrary initial position after $t$ MC steps per particle. The Lindemann parameter given by Eq.~\eqref{lindemann} measures the displacement of each particle relative to the average displacement of all particles that fall within a radius of the first minimum in $g(r)$ at $t=0$. Specifically, $\gamma_L (t \rightarrow \infty)$ diverges for a fluid but is bounded for a crystal or a glass. Following Zahn \emph{et al.} \cite{Zahn1999b}, we define the simulated system to be a fluid provided $\gamma_L (t_m)>0.033$, where $t_m$ is a large value for $t$ which we chose to be $t_m = 2000$. Once again, for two-component systems, $\gamma_L(t)$ is defined with respect to the large particles only.

We have checked for finite size effects in our MC simulations. We found that when we increased the number of particles from $N = 576$ to $N = 1089$ in the one component system, $\Gamma_c$ increased by less than 10\% for either the melting or freezing curves in Figure 1(a). This small shift lies within the uncertainty to which we determine $\Gamma_c$ and is therefore accurate enough for our purposes. We therefore conclude that a system size of $N = 576$ is large enough for obtaining an accurate measure of the freezing point for our dipolar systems.

\subsection{Density functional theory}\label{DFT}
In DFT, it is most convenient to study the system in the Grand Canonical Ensemble. For definiteness, let us consider the two-component colloidal system. In this case, the grand potential functional is given by
\begin{eqnarray}\label{gce}
  \Omega [\rho_1(\mathbf{r}), \rho_2(\mathbf{r})] &=& k_B T \sum_{i=1}^2 \int d\mathbf{r} \rho_i (\mathbf{r}) \left( \ln (\rho_i (\mathbf{r}) \Lambda_i^2)-1 \right) \nonumber \\
   &+& F_{ex}[\rho_1(\mathbf{r}), \rho_2(\mathbf{r})]+ \sum_{i=1}^2 \int d\mathbf{r} (\Phi_i(\mathbf{r})-\mu_i)\rho_i (\mathbf{r})
\end{eqnarray}
where $\rho_i(\mathbf{r})$ is the one-body density profile, $\Lambda_i$ is the (irrelevant) thermal wavelength, $\mu_i$ the chemical potential and $\Phi_i(\mathbf{r})$ the external potential acting on particles of species $i$ ($i=1,2$). In all the situations described below, we consider bulk systems, where $\Phi_i(\mathbf{r})=0$. The first term in Eq.~\eqref{gce} is the ideal gas (entropic) contribution to the free energy while the second term, $F_{ex}$, is the contribution from the interactions between particles and is called the excess Helmholtz free energy \cite{Hansenbook, Evans1979, evans1992fundamentals}.

The form of $F_{ex}$ is not known exactly for most systems, and the challenge in DFT is to construct accurate yet manageable approximations for this functional. Following Ramakrishnan and Yussouff \cite{Ramakrishnan1979}, we perform a series expansion of this functional about the homogeneous fluid state with uniform density $\rho_1, \rho_2$ up to second order in the density differences $\delta \rho_i(\mathbf{r})=\rho_i(\mathbf{r})-\rho_i$:
\begin{eqnarray}
  F_{ex}[\rho_1(\mathbf{r}), \rho_2(\mathbf{r})] &=& F_{ex}(\rho_1,\rho_2) + \sum_{i=1}^2 \int d \mathbf{r} \mu_{ex,i} \delta \rho_i (\mathbf{r}) \nonumber\\
   &-& \frac{k_BT}{2}\sum_{i,j} \int d\mathbf{r} \int d\mathbf{r}' \delta \rho_i (\mathbf{r}) c_{ij} (|\mathbf{r}-\mathbf{r}'|) \delta \rho_j (\mathbf{r}') \label{sot}
\end{eqnarray}
where $\mu_{ex,i}=\mu_i-k_BT\ln(\rho_i\Lambda_i^2)$ are the excess chemical potentials in the reference uniform liquid state and $c_{ij} (r)$ are the direct correlation functions calculated from the integral equation theory as described in subsection \ref{IET}. The key inputs that are required for our DFT are therefore the two-body direct correlation functions. Note that although formally the equilibrium density profile obtained from minimising the free energy functional corresponds to an ensemble average over all fluctuations in the system, in practice when making an approximation such as that in Eq.~\eqref{sot}, one is effectively neglecting certain fluctuation contributions. For further discussion on this issue, see e.g. ref.\cite{Evans1979, evans1992fundamentals, reguera2004role, archer2011interplay}. In particular, contributions from long-wavelength fluctuations are neglected, but for the freezing phenomena we consider here, these are less relevant.

Now van Teeffelen \emph{et al.} have shown that higher than second order terms in the series expansion of the excess Helmholtz free energy functional are required to accurately predict the freezing point $\Gamma_c$ for dipolar systems \cite{vanTeeffelen2006,vanTeeffelen2008}. However, in this paper we have opted to use the simpler free energy functional given by Eq.~\eqref{sot} where only terms up to second order are included. Although this means that we are no longer able to determine the freezing point accurately (such second order theories underestimate the stability of the crystal phase and therefore lead to predictions for $\Gamma_c$ that are too high \cite{vanTeeffelen2006,vanTeeffelen2008}), this simpler functional allows us to consider more complex dipolar systems with much greater ease compared to higher order theories. Specifically, we are able to consider both one- and two-component dipolar systems. We can also calculate the density profiles for the different particles \emph{ab initio}, without needing to make any \emph{a priori} assumptions concerning the form of these density profiles. Finally, we are able to perform calculations over large areas containing many unit cells where the system is not constrained by the boundary conditions to have a specific crystal structure but is free to choose its optimum structure.

One limitation of the Verlet closure that we have used in our DFT is the fact that the direct correlation functions $c_{ij}(r)$ can only be obtained for $\Gamma < \Gamma_c$. However, it is possible to extend our DFT to $\Gamma > \Gamma_c$ by using a heuristic scaling approximation for $c_{ij}(r)$ (see next section and Figure 3(a)). Although this approximation overestimates the stability of the crystal phase, it appears to generate crystal structures which are essentially the same as the equilibrium crystal structure (see Figure 3(b)) and therefore serve as good approximants for the equilibrium crystal structure. Furthermore, the resultant crystal structures remain (meta)stable for a small range of $\Gamma$ values below $\Gamma_c$, where our expressions for $c_{ij}(r)$ are accurate.

The equilibrium density profiles are those which minimise the grand potential Eq.~\eqref{gce}, i.e.\ which satisfy the following pair of Euler-Lagrange equations
\begin{equation}\label{EL-eq}
\frac{\delta\Omega[\rho_1(\mathbf{r}), \rho_2(\mathbf{r})]}{\delta\rho_i(\mathbf{r})}=0.
\end{equation}
Substituting Eqs.~\eqref{gce} and \eqref{sot} into Eq.~\eqref{EL-eq} we obtain
\begin{equation}\label{EL-eq2}
k_BT\ln\left(\rho_i(\mathbf{r})/\rho_i\right)+\sum_{i=1}^2  \int d\mathbf{r}' c_{ij} (|\mathbf{r}-\mathbf{r}'|) \delta \rho_j (\mathbf{r}')+\Phi_i(\mathbf{r})=0.
\end{equation}
We solve these equations using Piccard iteration. For more details on this method see e.g.\ Refs.~\cite{roth2010fundamental, hughes2014introduction}. There exist other more sophisticated approaches to minimising the functionals arising from DFT that can be used -- see e.g. ref.\cite{stopper2017massively} and references therein. We start from various initial guesses for the density profiles, including those obtained as a solution at slightly different parameter values (i.e.\ a neighboring state point) and also from uniform density profiles with small amplitude random noise fields added. In the second case, the system only goes to a crystalline solution with density peaks if the uniform liquid is linearly unstable \cite{Archer2014} and this is achieved by scaling the pair direct correlation functions $c_{ij}(r)$ to a higher value of $\Gamma$, as mentioned above. We also say more about this scaling procedure below.

\section{Results and Discussion}

\subsection{One-component system}

We start by considering the one-component system. Our first task is to determine the crystallization point $\Gamma_c$ for this system. As discussed in subsection \ref{MC}, $\Gamma_c$ was determined using MC simulations by measuring both the orientational order parameter and the Lindemann parameter as a function of $\Gamma$. In Figure 1(a), we plot the orientational order parameter $\Psi_6$ as a function of $\Gamma$, starting either from the fluid state (freezing curve) or from a perfect hexagonal crystal (melting curve). Both curves show a clear first order phase transition between $\Gamma = 11$ and $12$; the  minimal hysteresis between the two curves suggest that this range of $\Gamma$ values is close to the equilibrium freezing point. On the other hand, in Figure 1(b), we plot the dynamical Lindemann parameter $\gamma_L(t)$ as a function of $t$ (the number of MC steps per particle) for different values of $\Gamma$. In this case, there is a clear transition in the long `time' behaviour of $\gamma_L(t)$ at a slightly higher value of $\Gamma$, between $\Gamma=12$ and $13$, with $\gamma_L(t)$ diverging for $\Gamma \leq 12$ but converging to a finite value $\lesssim 0.033$ (indicated by the horizontal dashed line) for $\Gamma \geq 13$. This dynamical transition is also clearly seen in Figure 1(c) where we plot the long time value of the Lindemann parameter $\gamma_L (t_m=2000)$ as a function of $\Gamma$. The slight difference in the freezing point obtained from $\Psi_6$ and $\gamma_L(t)$ is not surprising given that they represent qualitatively different measures of the phase transition. In order to ensure that the system is in the fluid phase at the crystallization point, we define the crystallization point of the one-component system to be at the lower bound value of $\Gamma_c \approx 11$.

\begin{figure}
  \includegraphics[width=1.0\columnwidth]{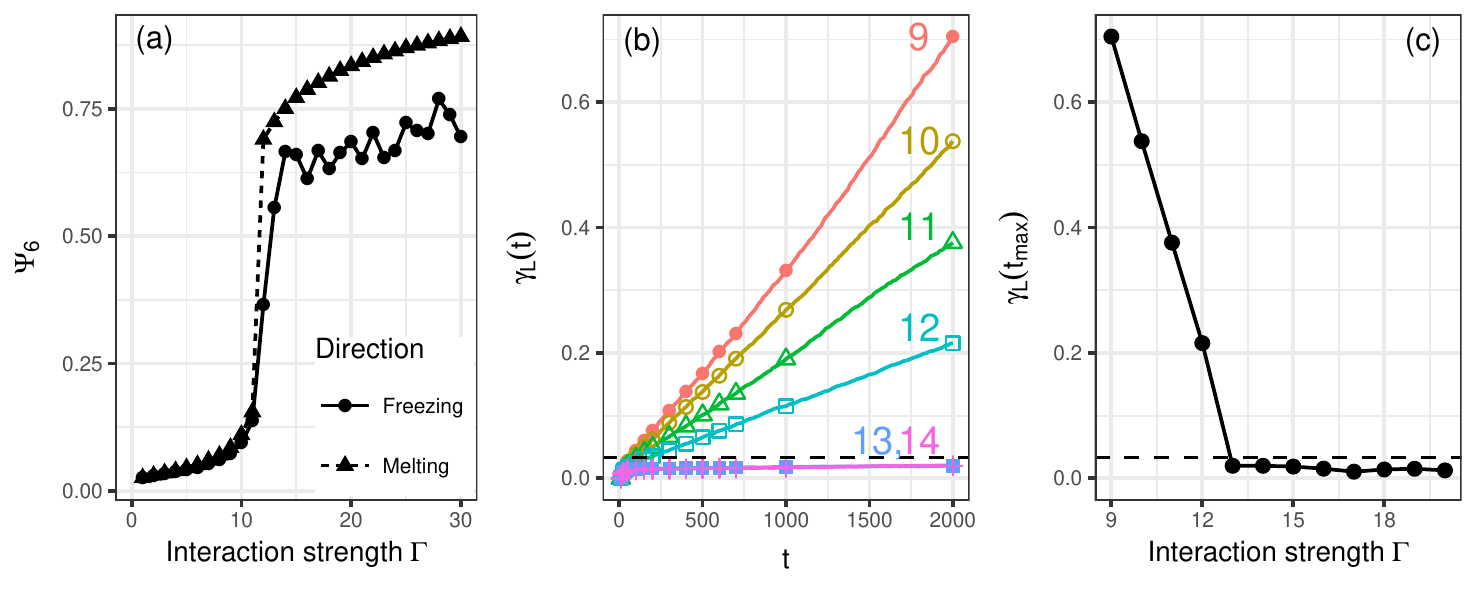}\\
  \caption{MC simulation results for determining the crystallization point of one-component dipolar monolayers: (a) Six-fold orientational order parameter $\Psi_6$ as a function of $\Gamma$, starting either from the fluid state (freezing curve) or from a perfect hexagonal crystal (melting curve); (b) Dynamical Lindemann parameter $\gamma_L(t)$ ($t$ is the number of MC steps per particle) for different values of $\Gamma$ ($\Gamma$ values labelled on each curve); (c) Long time value of the Lindemann parameter $\gamma_L (t_m=2000)$ as a function of $\Gamma$. The dashed horizontal line in (b), (c) corresponds to $\gamma_L = 0.033$, the threshold value of the Lindemann parameter in the crystal state.}\label{fig1}
\end{figure}

We next calculate the different correlation functions in the fluid phase (particularly $c(r)$) close to the crystallization point $\Gamma_c$ by solving the OZ relation and Verlet-HNC closure (see subsection \ref{IET}). In Figure 2(a), we plot the total correlation function $h(r)$ for $\Gamma = 11$ obtained from Verlet-HNC (solid line) compared to MC simulations (data points). The Verlet-HNC results are fairly insensitive to the choice of the cut-off length $r_c$ used in the closure \eqref{verlet3}, provided that $r_c$ is large enough. In all our calculations for both one and two-component systems, we choose $r_c = 9a$ (i.e., vertical dashed lines in Figure 2). We note that this value for $r_c$ is slightly smaller than the maximum value we could have chosen, i.e., half the MC simulation box size $L/2 = 12a$, but is larger than half the DFT calculation box (see Figure 4) and is therefore large enough for our purposes. For selected systems, we have also checked that using a larger value of $r_c$ (close to $L/2$) does not change our results for the liquid or crystal state. We see from Figure 2(a) that the agreement between Verlet-HNC and the MC simulations is very good, not just for $r \leq r_c$, but significantly also for $r > r_c$. These results show that the Verlet-HNC closure provides a very accurate description of the fluid structure across the entire range of $r$ values.

\begin{figure}
  \includegraphics[width=0.8\columnwidth]{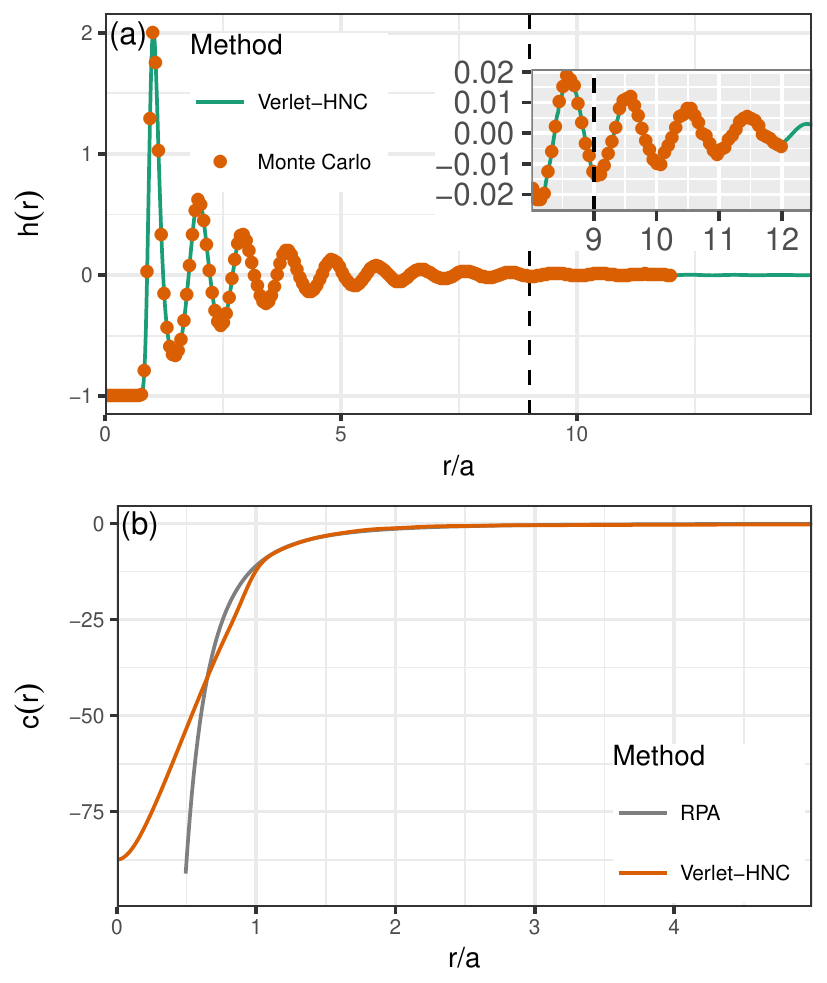}\\
  \caption{Liquid state correlation functions for one-component dipolar monolayers at $\Gamma=\Gamma_c=11$: (a) Total correlation function $h(r)$ obtained from the Verlet-HNC closure (solid line) and MC simulations (data points). The vertical dashed line is the cut-off length $r_c$ used in the closure \eqref{verlet2}. (b) Direct correlation function $c(r)$ obtained from the Verlet-HNC closure (red curve) and the random phase approximation (RPA, gray curve). }\label{fig2}
\end{figure}

In Figure 2(b), we plot the direct correlation function $c(r)$ for $\Gamma = 11$ obtained from the Verlet-HNC closure (red curve). We see that $c(r)$ is relatively featureless compared to $h(r)$, a fact that is well known from liquid state theory. However, it is instructive to compare the Verlet-HNC result with the much simpler random phase approximation (RPA) result $c_{RPA}(r)=-\beta U(r) =-\Gamma a^3/r^3$ \cite{Hansenbook} (gray curve). We see that, on the scale of the figure, there is good agreement between the two except at small $r$ ($r/a \lesssim 1$) where the RPA result diverges for $r \rightarrow 0$ but the Verlet-HNC result tends towards a large but finite negative value. The good agreement between the two curves at large $r$ suggests that the Verlet-HNC direct correlation function scales linearly with $\Gamma$ at large $r$, just like the RPA. To check whether this scaling also holds at small $r$, in Figure 3(a) we plot the Verlet-HNC results for $c(r)$ for a range of $\Gamma$ values up to $\Gamma=11$, while in the inset, we plot $c(r)/\Gamma$. The excellent collapse of the different curves onto a universal curve in the inset confirms that the RPA scaling $c(r)\sim \Gamma$ also holds to a good approximation for Verlet-HNC at low $r$ (provided $\Gamma \gtrsim 5$). However, the collapse of the different curves is in fact not perfect. For example, there is a small dispersion between the different curves around $r=0$ in the inset of Figure 3(a) which is hardly visible on the scale of the graph. We conclude from Figure 3(a) that the scaling approximation $c(r)\sim \Gamma$ preserves the essential features of the fluid structure, though it misses some subtle features of the structure.

\begin{figure}[th!]
  \includegraphics[width=0.8\columnwidth]{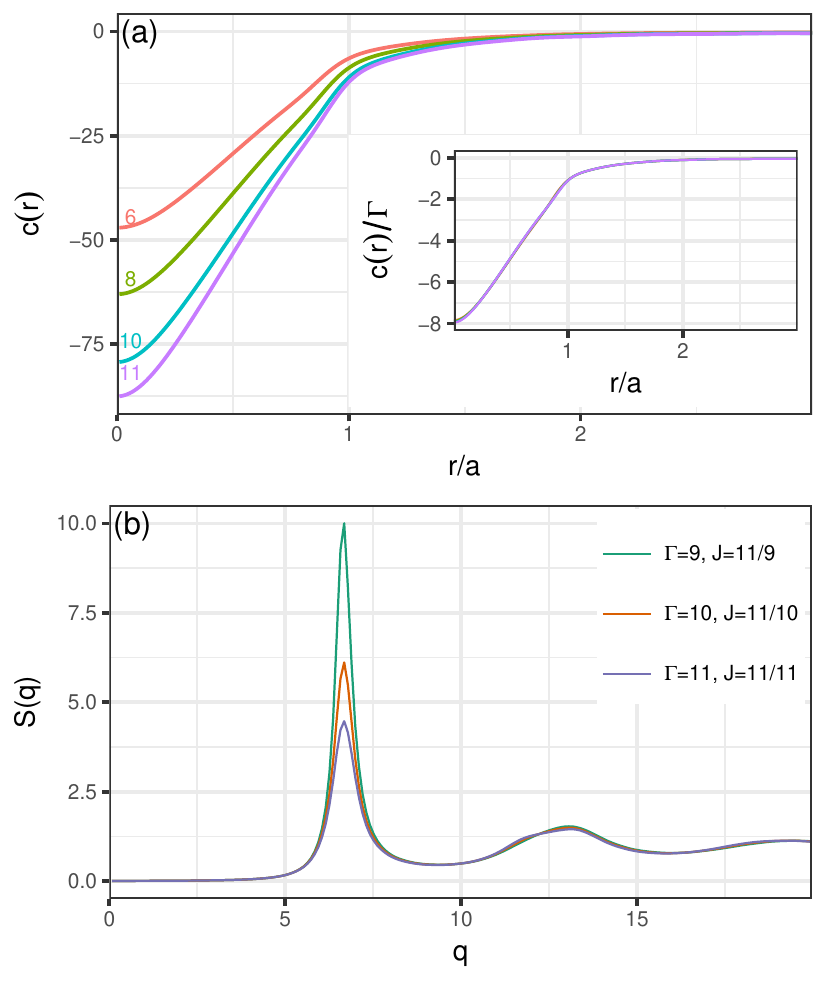}\\
  \caption{Checking the scaling approximation $c(r)\sim \Gamma$ for one-component dipolar monolayers: (a) Verlet-HNC results for $c(r)$ and $c(r)/\Gamma$ (inset) for a range of $\Gamma$ values up to $\Gamma=11$ ($\Gamma$ values labelled on each curve in main figure). Note the excellent collapse of the different curves in the inset; (b) Structure factor $S(q)$ calculated from the Verlet-HNC closure at $\Gamma = 11$ (purple curve), at $\Gamma=10$ with scale factor $J=11/10$ (orange curve) and at $\Gamma=9$ with scale factor $J=11/9$ (green curve).}\label{fig3}
\end{figure}

In order to probe in more detail what effect scaling $c(r)$ has on fluid structure, in Figure 3(b) we plot the structure factor $S(q)$ of the dipolar monolayer, calculated from the Verlet-HNC closure at $\Gamma = 11$ (purple curve), but also from the Verlet-HNC closure at lower values of $\Gamma$ ($\Gamma=9,10$, green and orange curve respectively) which are scaled to $\Gamma = 11$ using scale factors of $J=11/10, 11/9$ respectively ($J=\Gamma_{\textrm{target}}/\Gamma_{\textrm{original}}$, where $\Gamma_{\textrm{target}}$, $\Gamma_{\textrm{original}}$ are the target and original $\Gamma$ values respectively). We see that scaling $c(r)$ preserves the peak positions in the structure factor, but exaggerates the primary peak height, with the peak becoming increasingly prominent as we increase $J$. Indeed for even larger values of $J$ (e.g., scaling $c(r)$ from $\Gamma = 8$ to $\Gamma=11$ using $J=11/8$), the principal peak in $S(q)$ diverges, indicating that the fluid phase becomes linearly unstable and undergoes crystallisation \cite{Archer2014}. We conclude therefore that the scaling approximation for $c(r)$ underestimates the stability of the liquid phase. The approximation is nevertheless very useful as it generates crystal structures with the same Bragg peak positions (i.e., same symmetry) as the equilibrium crystal structure (Figure 3(b)). It also serves as a useful method for inducing the system to crystallise within our DFT calculation as we shall now demonstrate.

\begin{figure}
  \includegraphics[width=0.8\columnwidth]{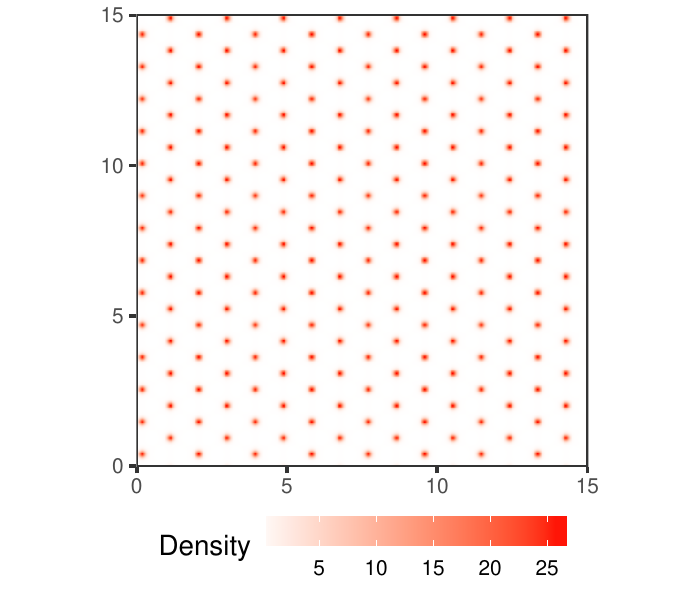}\\
  \caption{DFT results for the density profile of one-component dipolar colloidal monolayers at $\Gamma = 11$, showing the whole simulation box. Length scales in units of $a$ are indicated at the edges of the box.}\label{fig4}
\end{figure}

Having used integral equation theory to obtain accurate results for $c(r)$ close to the crystallization point, we now feed this information into our DFT to calculate the crystal structure. Using the uniform density fluid state (superposed with random noise) as our initial guess when solving the Euler-Lagrange equations \eqref{EL-eq2} at $\Gamma = \Gamma_c = 11$, we find that the fluid state is linearly stable. This is not surprising since, as discussed earlier, the effective crystallization point in our second order DFT is much higher than the equilibrium crystallization point. However, if we instead scale $c(r)$ at $\Gamma = 11$ with a scale factor of $J=1.3$ and feed this into our DFT, the uniform density state becomes linearly unstable and the system crystallizes into a hexagonal crystal. If we now use this crystal density profile as our initial guess when solving the Euler-Lagrange equations \eqref{EL-eq2} for a lower value of $J$, by gradually reducing $J$, following the crystalline solution branch until $J=1$ (i.e., no scaling), we find that we are able to obtain a hexagonal crystal state as a linearly stable solution at $\Gamma = 11$, as shown in Figure 4. Note that the grand potential for this crystal is higher than that of the fluid, indicating that the crystal is in fact metastable. The crystal remains linearly stable at $\Gamma = 10$ but melts at $\Gamma = 9$, showing that the \emph{spinodal} point for melting of the crystal structure lies between $\Gamma = 9$ and $\Gamma = 10$ in our DFT. Note that we are able to obtain the hexagonal crystal in Figure 4 without needing to make any \emph{a priori} assumptions about the form of the density profile or the symmetry of the crystal state. The DFT therefore provides a powerful \emph{predictive} tool for studying the crystal structure of dipolar monolayers.

\subsection{Two-component system}

We next turn our attention to the much richer case of two-component systems. We first consider the state point $\rho_2/\rho_1=2$ and $m_2=0.025$ where the dipole moment of the small particles is small enough to only slightly perturb the structure of the large particles. This state point has been studied by us previously \cite{Law2011,Law2011b} but our focus in those studies was on the crystallization of the \emph{small} particles. Here our focus is on the crystallization of the \emph{large} particles. We therefore determine the crystallization point of this two-component system $\Gamma_c$ by measuring the orientational order parameter and Lindemann parameter of the large particles only as a function of $\Gamma$ in our MC simulations.

\begin{figure}[t]
  \includegraphics[width=1.0\columnwidth]{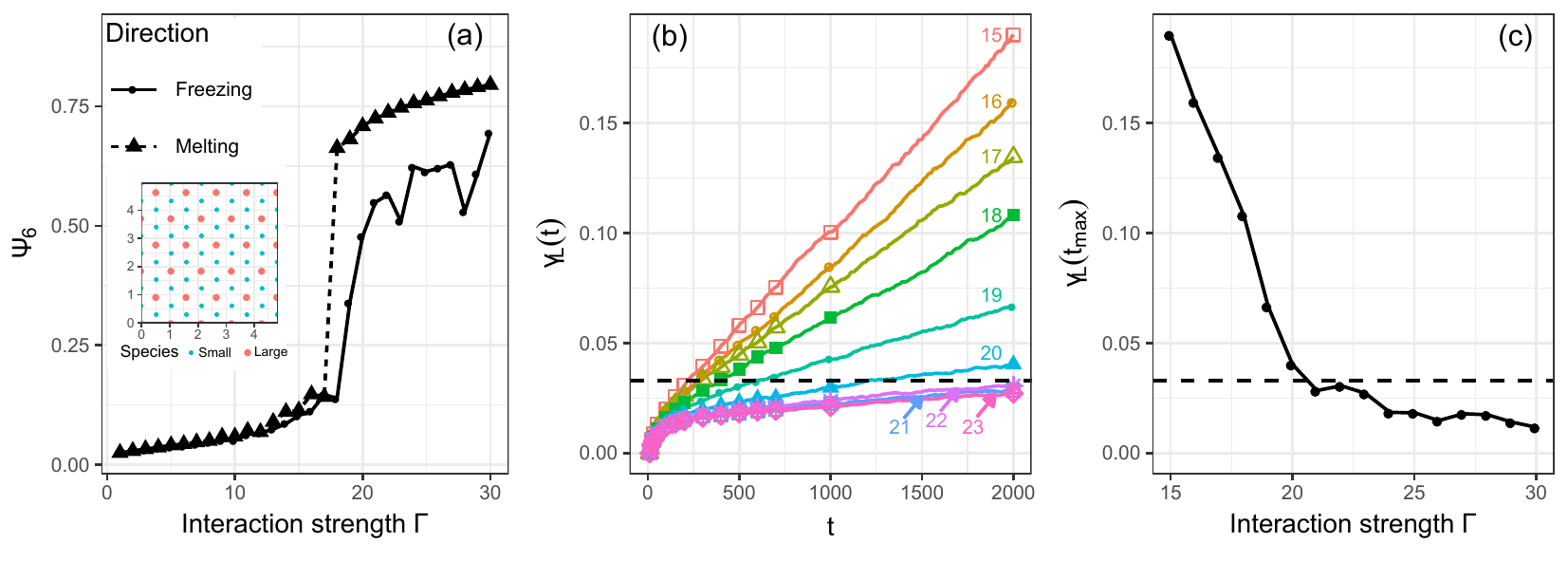}\\
  \caption{MC simulation results for determining the crystallization point of two-component dipolar monolayers with $\rho_2/\rho_1=2$ and $m_2=0.025$: (a) Six-fold orientational order parameter of large particles $\Psi_6$ as a function of $\Gamma$, starting either from the fluid state (freezing curve) or from a perfect hexagonal AB$_2$ crystal (melting curve). The inset shows the perfect hexagonal AB$_2$ crystal used as the initial state for the melting curves; (b) Dynamical Lindemann parameter for large particles $\gamma_L(t)$ ($t$ is the number of MC steps per particle) for different values of $\Gamma$ ($\Gamma$ values labelled on each curve); (c) Long time value of the Lindemann parameter for large particles $\gamma_L (t_m=2000)$ as a function of $\Gamma$. The dashed horizontal line in (b), (c) corresponds to $\gamma_L = 0.033$, the threshold value of the Lindemann parameter in the crystal state.}\label{fig5}
\end{figure}

In Figure 5(a), we plot $\Psi_6$ for the large particles as a function of $\Gamma$, starting either from the fluid state (freezing curve) or from a perfect hexagonal AB$_2$ crystal (melting curve). The inset shows the perfect hexagonal AB$_2$ crystal used as the initial state for the melting curves. Both curves show a clear first order phase transition around $\Gamma \approx  17$; the  minimal hysteresis between the two curves suggests that this $\Gamma$ value is close to the equilibrium freezing point of the large particles. Interestingly, the crystallization point in the two-component system occurs at a higher value of $\Gamma$ compared to that of the one-component system. This may be because the small particles in the two-component system remain disordered during the crystallization of the large particles (see Figure 8(d)), therefore introducing a higher degree of disorder in the structure of the large particles. In Figure 5(b) we plot $\gamma_L(t)$ for the large particles ($t$ is the number of MC steps per particle in the system) for different values of $\Gamma$, while in Figure 5(c) we plot $\gamma_L (t_m=2000)$ for the large particles as a function of $\Gamma$. We see that there is a clear transition in the long time dynamics around $\Gamma \approx 20$. In order to ensure that the system is in the fluid phase at the crystallization point for the liquid state structure calculations, we define the crystallization point for $\rho_2/\rho_1=2$ and $m_2=0.025$ to be at the lower bound value of $\Gamma_c \approx 17$.

\begin{figure}[t]
  \includegraphics[width=0.8\columnwidth]{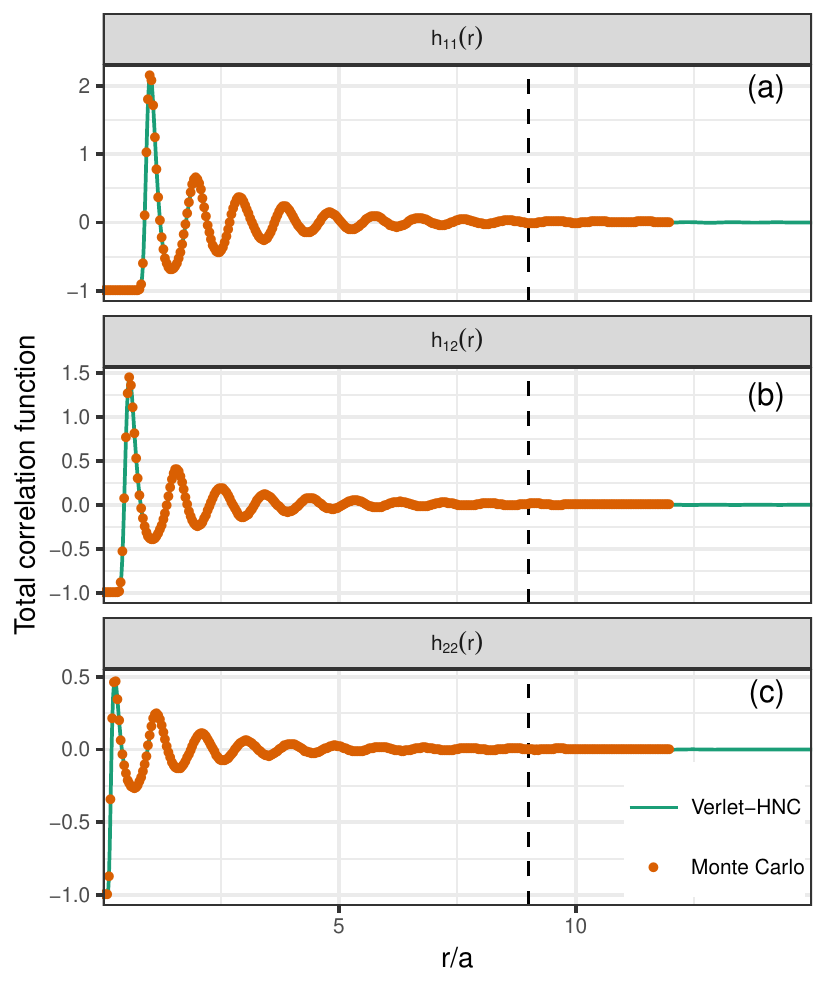}\\
  \caption{Total correlation functions for two-component system with $\rho_2/\rho_1=2$ and $m_2=0.025$ obtained from the Verlet-HNC closure (solid line) and MC simulations (data points): (a) $h_{11}(r)$ (b) $h_{12}(r)$ and (c) $h_{22}(r)$. The vertical dashed lines are the cut-off length $r_c$ used in the closure \eqref{verlet3}.}\label{fig6}
\end{figure}

We next use the Verlet-HNC closure to calculate correlation functions in the fluid phase close to the crystallization point. In Figure 6(a)-(c) respectively, we plot the total correlation functions $h_{11}(r)$, $h_{12}(r)$ and $h_{22}(r)$ respectively for $\Gamma = 17$, $\rho_2/\rho_1=2$ and $m_2=0.025$ obtained from the Verlet-HNC closure (solid line) compared to MC simulations (data points). The vertical dashed line represents the cut-off length $r_c = 9a$ that we used in Eq.~\eqref{verlet3}. The agreement between Verlet-HNC and the MC simulations is very good for both $r \leq r_c$ and $r > r_c$, indicating that the Verlet-HNC closure provides a very accurate description of the fluid structure for the two-component system across the entire range of $r$. In order to check whether the scaling approximation for the direct correlation function still holds for the two-component system, in Figure 7(a)-(c), we plot the Verlet-HNC results for the direct correlation functions $c_{ij}(r)$ ($i,j=1,2$) while in the insets we plot $c_{ij}(r)/\Gamma$ for $\rho_2/\rho_1=2$, $m_2=0.025$ and a range of $\Gamma$ values up to $\Gamma=17$. We see that there is good collapse of the different $c_{ij}(r)$ curves onto universal curves in the insets, apart from a small dispersion between the different curves around $r=0$. Interestingly, the dispersion is greater for $c_{12}(r)$ and $c_{22}(r)$ compared to $c_{11}(r)$. However, apart from this small discrepancy around $r=0$, we conclude that the scaling $c_{ij}(r)\sim \Gamma$ holds to a good approximation for two-component systems, provided $\Gamma$ is large enough.

\begin{figure}[t]
  \includegraphics[width=1.0\columnwidth]{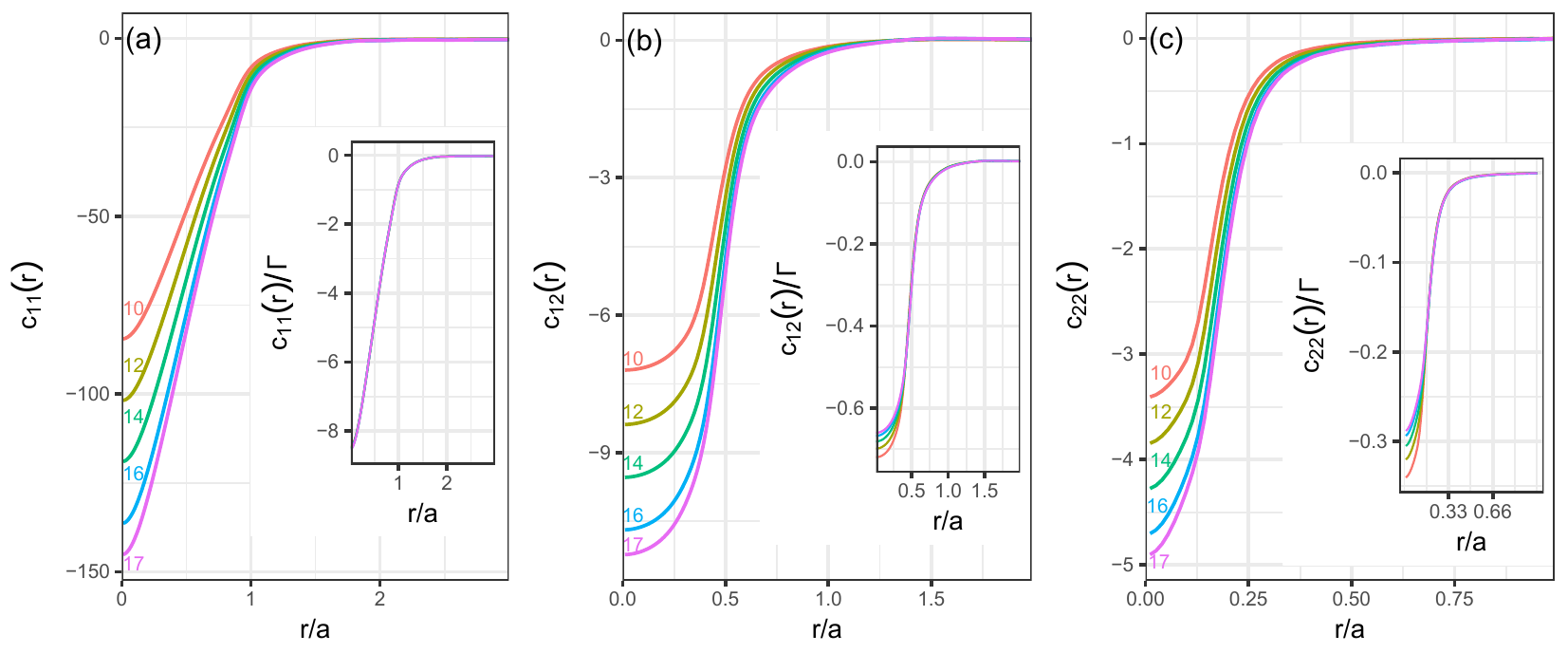}\\
  \caption{Direct correlation functions $c_{ij}(r)$ and $c_{ij}(r)/\Gamma$ (inset) for two-component system with $\rho_2/\rho_1=2$ and $m_2=0.025$ obtained from the Verlet-HNC closure for a range of $\Gamma$ values up to $\Gamma=17$: (a) $c_{11}(r)$; (b) $c_{12}(r)$; (c) $c_{22}(r)$.}\label{fig7}
\end{figure}

Having obtained accurate results for $c_{ij}(r)$ close to the crystallization point, we now feed this information into our DFT to calculate the crystal structure for the two-component system. In Figure 8(a)-(c), we show the crystal structure predicted by our DFT at $\Gamma=17$. The large and small particle density profiles shown are calculated as follows. We first scaled $c_{ij}(r)$ at $\Gamma = 17$ using a scale factor of $J=1.5$ and fed this into our DFT. This caused the large particles crystallize into a hexagonal crystal while the small particles remained delocalised in an interconnected honeycomb network around the large particles; the resultant structure is similar to the final structure shown in Figure 8(a)-(c). These profiles for $J=1.5$ are then used as the initial guess when solving the Euler-Lagrange equations for a lower value of $J$. By gradually reducing $J$ until $J=1$ (i.e., no scaling), we found that we are able to obtain the hexagonal crystal shown in Figure 8(a)-(c) as a linearly stable solution at $\Gamma=17$. We note that the grand potential for this crystal is higher than that of the fluid, indicating that the crystal is metastable. The large particle hexagonal lattice remains linearly stable at $\Gamma = 16$ but melts at $\Gamma = 15$, showing that the spinodal point for the melting of the large particle hexagonal lattice lies between $\Gamma = 15$ and $\Gamma = 16$ in our DFT.

\begin{figure}[t]
  \includegraphics[width=0.8\columnwidth]{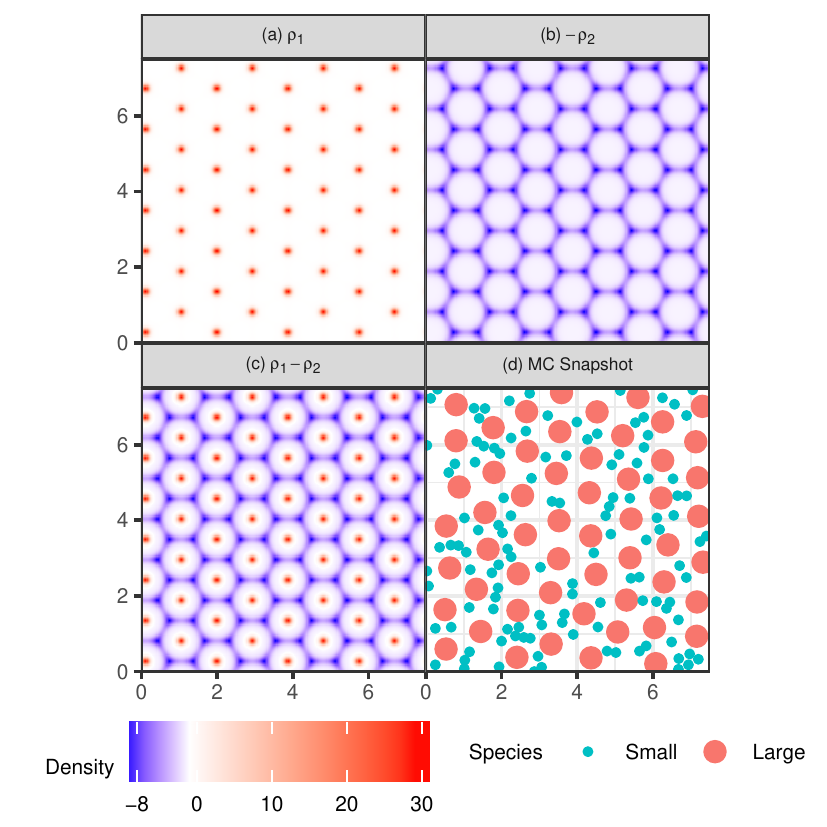}\\
  \caption{(a)-(c) DFT results for the density profiles of two-component system with $\rho_2/\rho_1=2$, $m_2=0.025$, $\Gamma=17$ for (a) large particle density profile $\rho_1(r)$; (b) (negative) small particle density profile $-\rho_2(r)$ and (c) Difference between the two $\rho_1(r)-\rho_2(r)$. In these plots, the large and small particle density profiles are represented by red and blue respectively. Only a quarter of the simulation box is shown in each case and length scales in units of $a$ are indicated at the edges of the box. (d) Snapshot from a MC simulation starting from the fluid state for the two-component system with $\rho_2/\rho_1=2$, $m_2=0.025$, $\Gamma=20$.}\label{fig8}
\end{figure}

From Figure 8(b), we note that there is a slight preference for the small particles to be at the interstitial sites between three large particles. However, the prominent density channels of small particles connecting these interstitial sites show that the small particles are in fact fluid, moving within the frozen lattice of large particles. These DFT predictions are confirmed by MC simulations. For example in Figure 8(d), we show a snapshot from a MC simulation starting from the fluid state for $\rho_2/\rho_1=2$, $m_2=0.025$ and $\Gamma = 20$ (i.e., $\Gamma \gtrsim \Gamma_c$), where we see that the large particles form a hexagonal lattice while the small particles are in a disordered fluid state. These results confirm our suggestion in Ref.~\cite{Law2011b} that for relatively small values of $m_2$, the melting of super-lattice structures for two-component systems proceeds via two distinct stages, corresponding to the melting of the small particle lattice at higher $\Gamma$ and the subsequent melting of the large particle lattice at a lower $\Gamma$.

Next we use our DFT to study the crystal structure for other state points. A comprehensive exploration of the parameter space of the two-component system lies beyond the scope of this paper. Here, we perform a preliminary exploration by just considering two nearby state points. Firstly, we consider the case $\rho_2/\rho_1 = 2$, $m_2 = 0.05$ in order to see what impact increasing $m_2$ has on the crystal structure. Secondly, we consider the case $\rho_2/\rho_1 = 1$, $m_2 = 0.2$ to see what impact changing the composition has on the crystal structure. Studying this state point also allows us to check if our DFT can produce crystal structures with symmetries other than the hexagonal symmetry since lattice sum calculations suggest that the zero-temperature structure of this state point is a square lattice \cite{Fornleitner2009}.

\begin{figure}[t]
  \includegraphics[width=1.0\columnwidth]{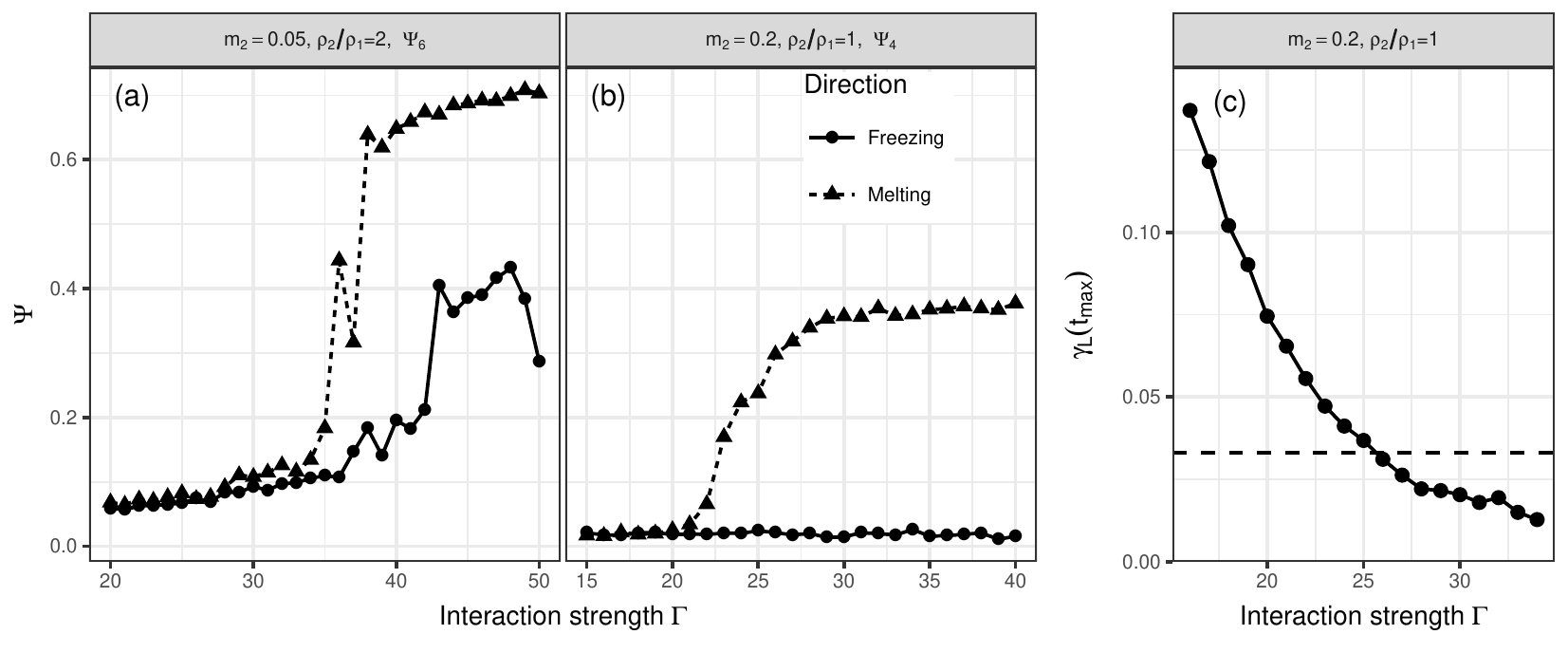}\\
  \caption{MC simulation results for determining the crystallization point of different two-component systems: (a) Six-fold orientational order parameter of large particles $\Psi_6$ for $\rho_2/\rho_1 = 2$, $m_2 = 0.05$ as a function of $\Gamma$, starting either from the fluid state (freezing curve) or from a perfect hexagonal AB$_2$ crystal (melting curve); (b) Four-fold orientational order parameter of large particles $\Psi_4$ for $\rho_2/\rho_1 = 1$, $m_2 = 0.2$ as a function of $\Gamma$, starting either from the fluid state (freezing curve) or from a perfect square AB crystal (melting curve); (c) Long time value of the Lindemann parameter for the large particles $\gamma_L (t_m=2000)$ for $\rho_2/\rho_1 = 1$, $m_2 = 0.2$ as a function of $\Gamma$. The dashed horizontal line corresponds to $\gamma_L = 0.033$.}\label{fig9}
\end{figure}

As before, we first determine the crystallization point $\Gamma_c$ for these state points. In Figure 9(a), we plot $\Psi_6$ for large particles for $\rho_2/\rho_1 = 2$, $m_2 = 0.05$ as a function of $\Gamma$ from MC simulations, starting either from the fluid state (freezing curve, circles) or from a perfect hexagonal AB$_2$ crystal (melting curve, triangles). The melting curve shows a clear first order phase transition around $\Gamma \approx  35$. The freezing curve exhibits a first order phase transition at a significantly higher $\Gamma$ value and the transition is also less clear cut. In addition, the value of $\Psi_6$ in the crystal state is significantly lower for the freezing curve compared to the melting curve for this state point. This is in contrast to the case $\rho_2/\rho_1=2$, $m_2=0.025$ where the discrepancy in the value of $\Psi_6$ for the crystal state between the freezing and melting curves is much smaller (see Figure 5(a)). These results suggest that increasing $m_2$ lowers the $\Gamma$ value at which the glass transition occurs, thus preventing the system from achieving full crystalline order at higher values of $\Gamma$ because the simulations are close to a glassy state. In order to ensure that the system is in the fluid phase at the crystallization point, we define the crystallization point for $m_2 = 0.05$, $\rho_2/\rho_1 = 2$ to be at the lower bound value of $\Gamma_c \approx 35$.

The influence of $m_2$ on the glass transition is confirmed in Figure 9(b) where we plot $\Psi_4$ for large particles for the state point $\rho_2/\rho_1 = 1$, $m_2 = 0.2$ as a function of $\Gamma$ from MC simulations, starting either from the fluid state (freezing curve, circles) or from a perfect square AB crystal (melting curve, triangles; see Figure 10(b) for an example of the square AB lattice). The reason why we measure $\Psi_4$ and use a square lattice as the starting point for the melting curve is because the zero temperature structure for this state point is the square lattice, as mentioned earlier. Clearly, the first order crystallization transition is completely absent from the freezing curve, presumably because crystallization has been arrested by a glass transition for such a large value of $m_2$. The results in Figure 9(b) are consistent with experiments on binary dipolar systems which show that for $m_2 \approx 0.1$, the fluid phase is arrested by a glass transition before crystallization can occur \cite{Ebert2008}. The aim of this paper is to construct a \emph{predictive} model for the structure of binary dipolar systems and we therefore require a route for measuring $\Gamma_c$ that does not rely on any \emph{\emph{a priori}} knowledge of the final crystal structure. Since this is not possible using the orientational order parameter route in this case, we use the Lindemann parameter route to measure $\Gamma_c$ instead. From Figure 9(c), the Lindemann parameter method yields $\Gamma_c \approx 25$ for $\rho_2/\rho_1 = 1$, $m_2 = 0.2$ (here $\Gamma_c$ refers to the point where the system ceases to be a fluid either through crystallization or a glass transition). Interestingly, unlike the previous two-component state points studied in this paper, there is no clear transition in the long time Lindemann parameter curve at $\Gamma_c$, consistent with the fact that it is a glass transition.

Having determined $\Gamma_c$ for the different cases considered above, we next use Verlet-HNC to calculate the direct correlation functions and use these in our DFT to calculate the resultant crystal structure. In Figure 10(a), we show the crystal structure predicted by our DFT for $\rho_2/\rho_1 = 2$, $m_2 = 0.05$ and $\Gamma=\Gamma_c =35$. The density profiles shown are calculated as follows: We first scaled $c_{ij}(r)$ at $\Gamma = 35$ using a scale factor of $J=1.2$ and fed this into our DFT. This caused the large particles to crystallize into a hexagonal crystal while the small particles remained delocalised in an interconnected honeycomb network around the large particles; the resultant structure is similar to the final structure shown in Figure 10(a). These profiles for $J=1.2$ are then used as the initial guess when solving the Euler-Lagrange equations for a lower value of $J$. By gradually reducing $J$ and following the crystalline solution branch until $J=1$ (i.e., no scaling), we find that we are able to obtain the hexagonal crystal state shown in Figure 10(a) as a linearly stable solution at $\Gamma=35$. We note that the grand potential for this crystal is higher than that of the fluid, indicating that the crystal is metastable.

\begin{figure}[th!]
  \includegraphics[width=0.6\columnwidth]{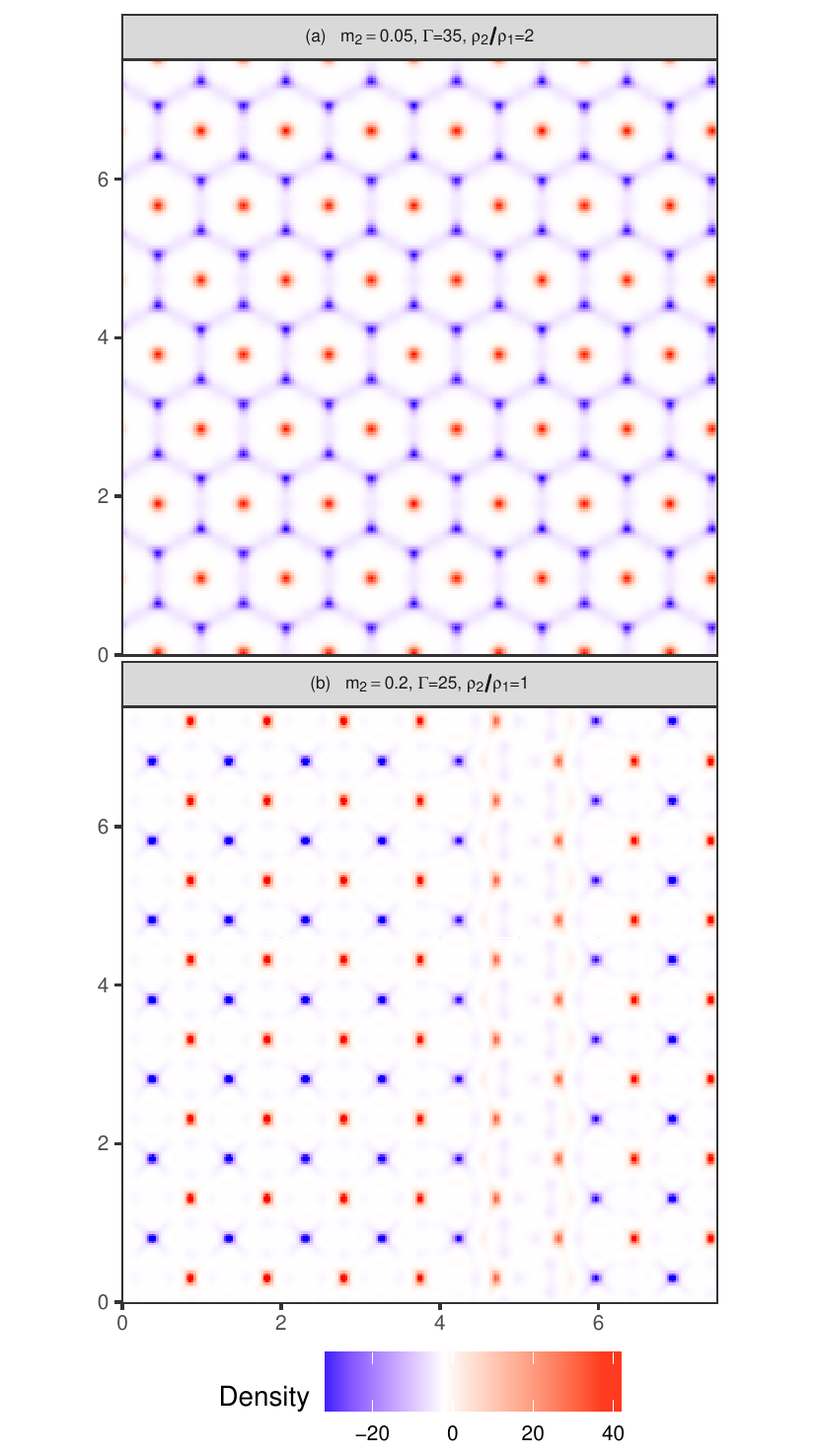}\\
  \caption{DFT results for the difference in the density profiles of large and small particles $\rho_1-\rho_2$ for different two-component systems: (a) $\rho_2/\rho_1 = 2$, $m_2 = 0.05$ and $\Gamma=\Gamma_c =35$; (b) $\rho_2/\rho_1 = 1$, $m_2 = 0.2$, $\Gamma=\Gamma_c =25$. In these plots, the large and small particle profiles are represented by red and blue respectively. Only a quarter of the simulation box is shown in each case and length scales in units of $a$ are indicated at the edges of the box. Note in (b) the line dislocation along the line $x\approx5$. }\label{fig10}
\end{figure}

The crystal structure in Figure 10(a) (for $\rho_2/\rho_1 = 2$, $m_2 = 0.05$) is similar to the one in Figure 8(c) (for $\rho_2/\rho_1 = 2$, $m_2 = 0.025$). However, we note that the small particles in Figure 10(a) are more localised at the interstitial sites between three large particles compared to Figure 8(c). This is not surprising since the larger dipole moment of the small particles in Figure 10(a) means that they effectively experience a larger confining potential from the large particles. We also note that for $\rho_2/\rho_1 = 2$, $m_2 = 0.05$, lattice sum calculations of the zero-temperature structure (i.e., $\Gamma \rightarrow \infty$) show that the dipole moment of the small particles is large enough to distort the hexagonal lattice of the large particles \cite{Fornleitner2009,Law2011b}. However, Figure 10(a) shows no such distortion. We conjecture that this is because Figure 10(a) shows the structure of the system at finite $\Gamma$ where the small particle lattice has melted while the large particle lattice remains intact. The delocalisation of the small particles leads to the small particles exerting a much smaller net force on the large particles due to averaging, such that the large particle lattice is no longer distorted.

In Figure 10(b), we show the crystal structure predicted by our DFT for $\rho_2/\rho_1 = 1$, $m_2 = 0.2$, $\Gamma=\Gamma_c =25$. The calculation of the crystal structure for this state point was more involved compared to the previous ones considered in this paper. We found that it was not possible to induce the system to crystallize by scaling $c_{ij}(r)$ at $\Gamma = 25$. Instead, we needed to scale $c_{ij}(r)$ at a lower $\Gamma$ ($\Gamma=17$) using a large scale factor ($J=1.9$). This is not surprising since $\Gamma_c$ represents the glass transition point (rather than crystallization point) in this case, where the system is still very far from crystallization. Stronger scaling of the direct correlation functions is therefore required to induce crystallization, which is provided by using direct correlation functions at a lower $\Gamma$ in conjunction with a larger scale factor (see Figure 3(b)). In addition, using a uniform density profile (superposed with random noise) as our initial guess when solving the Euler-Lagrange equations produced a crystal structure with a large number of defects. To overcome this problem, we used a step function density along one edge of the simulation box (superposed with random noise) as our initial density profile instead (c.f.\ Ref.\cite{Archer2014}). This procedure greatly reduced the number of defects, though a line dislocation is still discernable in the final structure shown in Figure 10(b). Finally, the profile obtained from $\Gamma=17$, $J=1.9$ was used as the initial guess when solving the Euler-Lagrange equations for $\Gamma=25$, $J=1.9$. The scale factor $J$ was then reduced very slowly, allowing us to obtain the crystal state shown in Figure 10(b) as a linearly stable solution at $\Gamma=25$ with no scaling.

From Figure 10(b) we see that our DFT predicts a square superlattice structure rather than a hexagonal structure for $\rho_2/\rho_1 = 1$, $m_2 = 0.2$, $\Gamma=25$. Our DFT also predicts that the small particles are localised, unlike the previous state points studied in this paper. The latter is not surprising given the much larger dipole moments of the small particles in this case. The square superlattice structure is in excellent agreement with lattice sum calculations of the zero temperature structure for this state point \cite{Fornleitner2009}. It is also consistent with experiments on binary dipolar systems for $m_2 \approx 0.1$ where square superlattice structures are found locally, though the experimental system is disordered globally as it is trapped in a glassy state \cite{Ebert2008}. The results in Figure 10 illustrate that our DFT is capable of producing a variety of stable crystal structures from first principles.

For future work, it would be useful to perform a comprehensive exploration of the parameter space of the two-component system using our DFT. As our preceding discussion shows, the method works best when the computer simulation derived direct correlation functions are available close to the crystallization point (e.g., the case of Figure 10(a)). However, our DFT remains relevant even when the computer simulation data is limited by glass transitions to be far from the crystallization point (e.g., the case of Figure 10(b)), though more care is required in this case when solving the Euler-Lagrange equations to obtain stable crystal structures. We note that we did not observe any demixing of the two species in the two-component system for any of the parameters explored in this paper. This is consistent with previous results on two-component dipolar systems \cite{Hoffmann2006,Hoffmann2006b} and is a consequence of the fact that the non-additivity parameter for this system is negative for all dipole moment ratios \cite{Hoffmann2006,Hoffmann2006b}. Physically, a negative non-additivity parameter means that particles on average dislike the opposite species less than they dislike their own species, thus suppressing demixing of the different species. In principle, a negative non-additivity parameter does not preclude solid-solid phase separation occurring. However, we did not observe any evidence for such phase separations for the any of the system parameters we explored.

Finally, we point out that our DFT approach, where we couple a second order theory (SOT) with simulation-derived direct correlation functions, is in principle applicable to other types of interactions and is not restricted to dipolar interactions alone. However, as the SOT under-predicts the freezing temperature \cite{vanTeeffelen2006,vanTeeffelen2008} while the simulation-derived direct correlation functions are only available above the freezing temperature, we need an additional relation to extend the direct correlation function below the freezing temperature in order to study crystallization phenomena. For dipolar interactions, this extension is provided by our heuristic scaling approximation. However, it is a non-trivial problem to carry out this extension for a general potential and this is the key rate-limiting step that we need to overcome in order to apply our DFT approach to other types of interactions.

\section{Conclusions}

We have constructed a DFT for both one and two-component dipolar monolayers. Our theory utilises a series expansion for the excess Helmholtz free energy functional, truncated at second order in the density profile. Although this simplification means that we cannot determine the freezing point accurately, our approach allows us to calculate \emph{ab initio} both the density profile and symmetry of the final crystal structure for both one- and two-component dipolar systems. For experimentally realistic interactions such as dipolar interactions, we found that very accurate results for the direct correlation functions are required as input to the DFT, as more simple approximations often used for soft potentials such as the random phase approximation (RPA), or even the hypernetted chain closure (HNC) and the Rogers-Young closure (RY), are not accurate enough. We therefore employed direct correlation functions based on computer simulations which are accurate up to the crystallization point of the system. We also used a heuristic scaling approximation which allowed us to extrapolate the simulation-derived direct correlation functions beyond the crystallization point of the computer simulations and induce the system to crystallize within our DFT.

Our DFT predicts hexagonal crystal structures for one-component systems, and a variety of superlattice structures for two-component systems, including those with hexagonal and square symmetry. These predictions are in good agreement with known theoretical and experimental results for these systems. The theory also provides new insights into the structure of the two-component system in the intermediate temperature regime where the small particle lattice has melted but the large particle lattice remains intact. As such, the DFT bridges the gap between integral equation theory, which works well at high temperatures and lattice sum methods, valid at zero temperature, giving us a powerful \emph{predictive} tool for studying the crystallization of dipolar monolayers. For future work, it would be useful to perform a comprehensive exploration of the parameter space of the two-component system using our DFT. It would also be useful to extend the model by including third order terms in the free energy expansion \cite{vanTeeffelen2006,vanTeeffelen2008} in order to obtain more accurate results for both the freezing points and the free energies of the different crystal structures.

\emph{Acknowledgements.} We gratefully acknowledge financial support from the EPSRC for this work (Grant number EP/L025078/1 Self assembly of two dimensional colloidal alloys for metamaterials applications). We also acknowledge the Viper High Performance Computing facility of the University of Hull and its support team.

%\bibliography{DFT}

\end{document}